# Peering through SPHERE Images: A Glance at Contrast Limitations

Faustine Cantalloube[1]
Kjetil Dohlen[2]
Julien Milli[3]
Wolfgang Brandner[1]
Arthur Vigan[2]

[1] Max Planck Institute for Astronomy, Heidelberg, Germany
[2] Laboratoire d'Astrophysique de Marseille, Aix Marseille Université, CNRS, CNES, France
[3] ESO

Various structures are visible within Spectro-Polarimetric High-contrast Exoplanet REsearch instrument (SPHERE) images that are not always straightforward to interpret. In this article we present a review of these features and demonstrate their origin using simulations. We also identify which expected or unexpected features are limiting the contrast reached by the instrument and how they may be tackled. This vision paves the way to designing a future upgrade of the SPHERE instrument and the next generation of high-contrast instruments such as those planned for the Extremely Large Telescope (ELT).

## Imaging exoplanet and circumstellar discs with SPHERE

Direct imaging provides key information for understanding the nature of exoplanets, their formation and evolution processes. It is complementary to other detection techniques since (i) it is biased towards distant giant gaseous planets hosted by young stars, (ii) it enables direct extraction of the planet's thermal emission and its spectrum and (iii) it reveals the planet in its birth environment and its connection to the circumstellar disc's properties. Dedicated instruments have been built worldwide to scrutinise very short separations (below 500 milliarcseconds), and reach very high contrasts (more than $10^{-6}$) between the host star and its planetary companions in the near infrared (NIR).

In 2014, the SPHERE instrument was installed on Unit Telescope 3 (Melipal) of the Very Large Telescope (VLT) at the ESO Paranal Observatory (Beuzit et al., 2019). The common path infrastructure of SPHERE is equipped with an extreme adaptive optics (AO) system known as SAXO (Fusco et al., 2006) and coronagraphs (such as the apodised Lyot Coronagraph, APLC; see Carbillet et al., 2011 and Guerri et al., 2011), allowing it to recover the diffraction-limited angular resolution of the 8-metre telescope and to reach a contrast of $10^{-4}$ at a few hundred milliarcseconds in the raw images. This common path infrastructure feeds three scientific instruments[1]: the InfraRed Dual-band Imager and Spectrograph (IRDIS; Dohlen et al., 2008), the Integral Field Spectrograph (IFS; Claudi et al., 2008) and the Zurich Imaging POLarimeter (ZIMPOL; Schmid et al., 2018).

Since 2014, the SPHERE instrument has delivered a wide variety of astrophysical results and impressive images (including nine ESO press releases[2]). So far, SPHERE has contributed to the discovery of two confirmed exoplanets (HIP65426b, Chauvin et al., 2017; and PDS70b, Keppler et al., 2018) and additional candidates are being followed up with the SpHere INfrared survey for Exoplanets (SHINE; Chauvin et al., 2017) within the Guaranteed Time Observations (GTO). From the current results, a clear paucity of giant planets appears beyond typically 10 astronomical units. Can we improve the current SPHERE limitations to access closer separations and fainter objects?

The unique combination of extreme AO correction and the coronagraph reveals very faint structures in the images that were not always expected. In order to better understand the images delivered by SPHERE and to provide clues about its current limitations, and thus to shed light onto future high-contrast instrumentation pathways, we present and explain the various features visible in the SPHERE images (see Figure 1) by comparing on-sky images to simulations.

To simulate SPHERE images, the following are included: (i) the VLT pupil showing central obstruction and spiders holding the secondary mirror; (ii) von Karman atmospheric turbulence (at seeing ~ 0.85 arcseconds) and its correction by the SAXO AO; (iii) the Apodized Pupil Lyot Coronagraph (APLC); (iv) the remaining high-order optical aberrations within SPHERE; and (v) photon and detector noise. The overall light path is schematically presented in Figure 2 (from left to right) where the light is propagated from pupil planes to focal planes by a Fourier transform, under the Fraunhofer farfield approximation. At the entrance of this setup, the SAXO residual phase is simulated using a Fourier-based analytical code (Jolissaint et al., 2006) from which an instantaneous coronagraphic exposure is produced. "Pseudo-long exposures", are obtained by summing 500 temporally uncorrelated exposures.

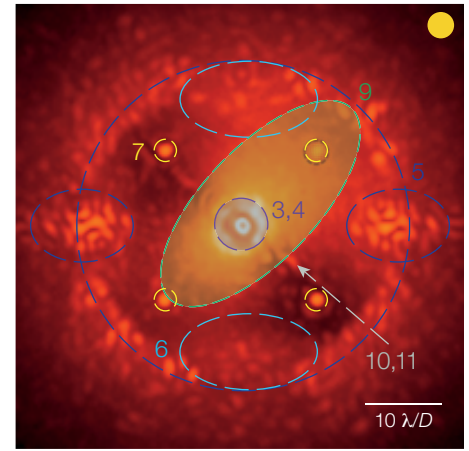

Figure 1. Typical image obtained with the SPHERE-IRDIS instrument (H2 narrow-band filter at 1.6 μm). The various structures are highlighted with numbers referring to the figure(s) in which they are explained. The yellow dot in the top right-hand corner indicates that this is a real on-sky image taken with SPHERE.

The final simulated images have the same spectral response as the SPHERE filters[3], by summing images at different wavelengths over every nanometre. Most of the figures presented here show two simulated images: one using an ideal coronagraph, which perfectly cancels the light diffracted by a fully circular telescope aperture (Sauvage et al., 2010), and which is affected only by residual phase errors from the AO system in addition to the phase term from which the feature originates; the other one using an APLC coronagraph, affected by all of the main phase error terms visible in SPHERE images. The figures illustrating the pupil plane images are shown in blue and the focal plane images are shown in red. A yellow circle in the top right-hand corner indicates real images from SPHERE. All images are cropped to 200 × 200 pixels (about 2.5 arcseconds).

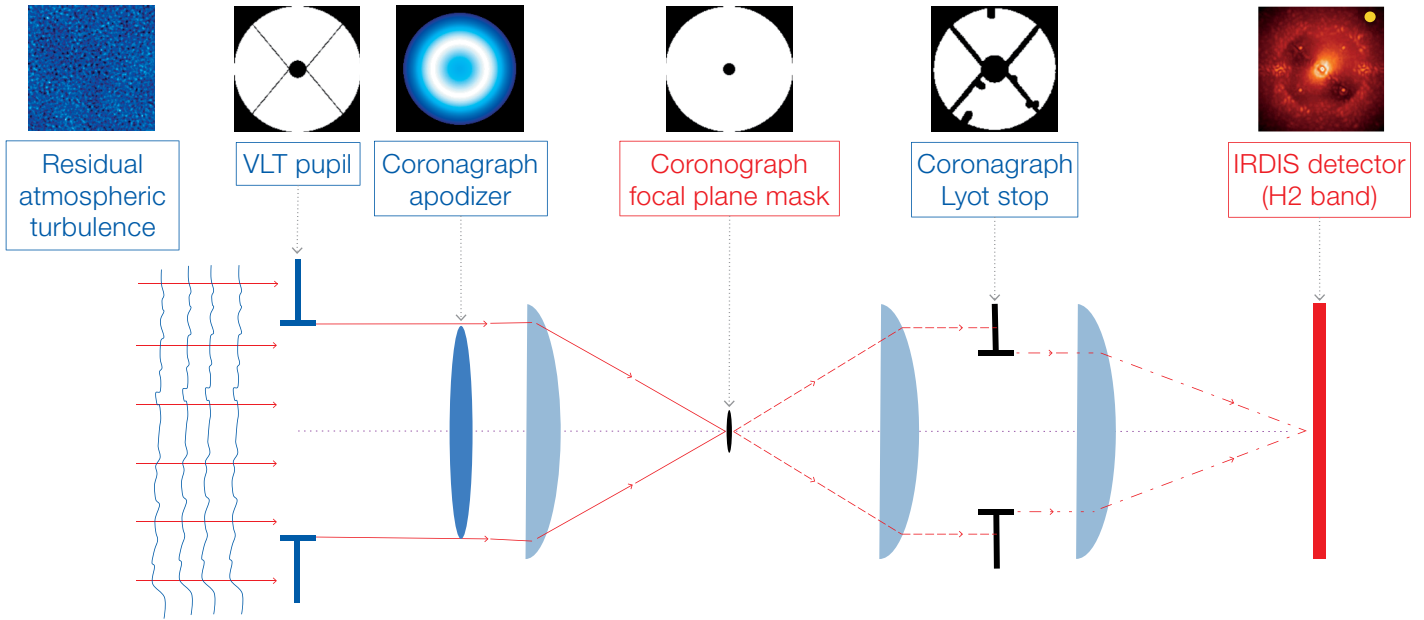

Figure 2. SPHERE-IRDIS APLC coronagraphic setup used to produce the simulations. Phase and amplitude are shown in the pupil planes (blue) and intensity in the detector plane (red).

### Dissection of a SPHERE image

The coronagraph is essential to reaching high contrast at close separation from the star. Its role is to suppress as much diffracted light from the star as possible while preserving any other astrophysical signal present in the field of view.

### Effect of the coronagraph (Figure 3)

Without a coronagraph, the diffracted light of the central star, i.e., the point spread function (PSF), hides its environment (Figure 3a). It is possible to block the light of the inner core of the PSF by placing an opaque mask in the focal plane before re-imaging the star (Figure 3b). The very central part of the image shows a bright spot where one might naïvely expect a dark spot; this is the so-called Poisson or Arago spot, which is due to diffraction by the coronagraphic focal plane mask.

To remove the diffraction pattern due to the telescope aperture, a Lyot stop is placed in the following pupil plane, consisting of a wider central obstruction, wider spiders and smaller outer diameter (Figure 3c). In order to smooth the sharp edges of the VLT pupil and thus avoid strong diffraction effects (ripples in the focal plane, as in the Gibbs effect), a pupil apodiser is placed upstream of the coronagraph focal plane mask (Figure 3d). Its transmission function has been optimised to avoid these ripples while keeping a high throughput and resolution (Soummer et al., 2005). As a result, under very good conditions SPHERE can reach a contrast of up to $10^{-4}$ at 250 milliarcseconds in the raw coronagraphic image with a 50% transmission at 100 milliarcseconds in the $H$-band[4].

### Effect of the VLT pupil on the coronagraph signature (Figure 4)

The coronagraph design, its resulting performance, and the central image pattern are all driven by the shape of the telescope pupil. From a circular pupil (Figure 4a) to a centrally obscured pupil (Figure 4b), a brighter ring appears close to the star. When the spiders are added this bright secondary ring is broken into four petals (Figures 4c and d). These patterns are strongly dependent upon the observing wavelength as the size of the focal plane mask is fixed — Figure 4 shows the specific case of Y-band images at 1.02 μm to highlight this effect.

Figure 3. Illustration of the APLC coronagraph effect ($H2$-band): (a) non-coronagraphic image, (b) coronagraphic image with only the focal plane mask, (c) adding the Lyot stop downstream of the focal plane mask, (d) adding the SPHERE pupil apodiser upstream of the focal plane mask.

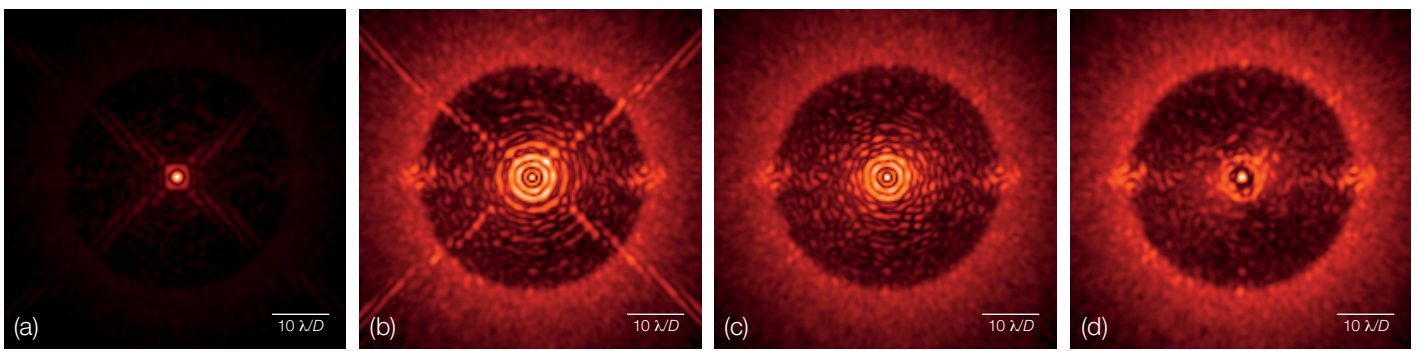

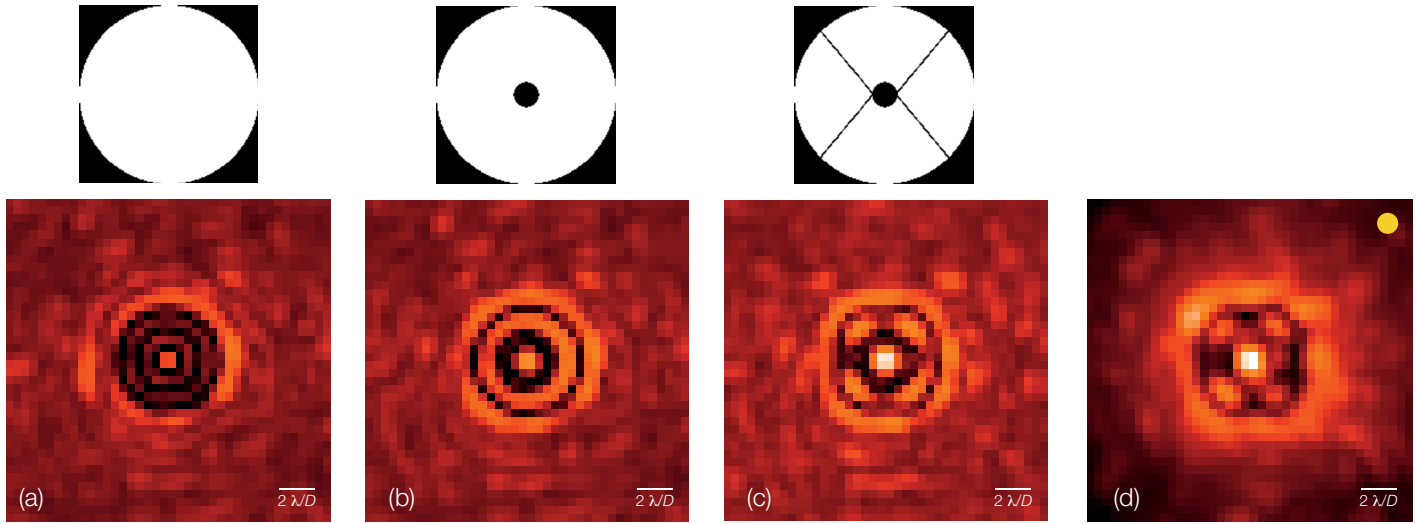

A second type of feature apparent on the images originates from the SAXO system. SAXO is composed of three main elements: (a) a piezo stack high-order deformable mirror (HODM) with 41 actuators across the pupil and a tip-tilt deformable mirror (TTDM) to modulate the incoming phase distorted by the atmospheric turbulence; (b) a Shack-Hartman (SH; Sauvage et al.,

2014) wavefront sensor (WFS) to sense the phase of the incoming wavefront at 1380 Hz; and (c) a real time computer to analyse the wavefront and compute the correction command to be sent to the deformable mirrors (DMs) in real time. When the target is faint, the main error comes from the measurement noise. In the following, the target star is considered bright enough (less than 8 magnitudes in the V-band) to ignore this error.

### The fitting error (Figure 5)

As the number of HODM actuators is finite, only the low frequencies up to the cut-off frequency, defined by the HODM inter-actuator pitch, can be corrected. The simulated AO residual phase including only the fitting error (Figure 5a) shows residuals with a typical size equal to the inter-actuator spacing and smaller. In the focal plane image it creates a central circular dark zone, called the corrected area, delimited by the correction ring

Figure 4. Illustration of the VLT pupil effect on the APLC coronagraphic images (Y2-band): (a) full circular aperture, (b) pupil with central obstruction, (c) VLT pupil including spiders and central obstruction and (d) corresponding on-sky SPHERE-IRDIS image.

(Figure 5b). The seeing-limited region lies outside of this corrected area, where the contrast reached is primarily limited by the seeing conditions.

In SPHERE images, the correction ring shows two bright patterns in the horizontal direction (Figure 1, dark blue) which are due to the imprint of the HODM actuator grid. The HODM is made up of linear arrays of 22 piezostack actuators joined in the middle. We can visualise this HODM grid by using the Zernike sensor for Extremely accurate measurements of Low-level Differential Aberrations (ZELDA; N' Diaye et al., 2013) that is a phase mask placed at the location of the coronagraph FPM converting upstream phase errors into intensity. An example

Figure 5. Illustration of the correction radius due to the fitting error (H2-band): (a) AO residual phase with only the fitting error, (b) corresponding ideal coronagraphic image showing a perfect dark hole, (c) the real HODM physical shape is not homogeneous and (d) this results in additional patterns in the image.

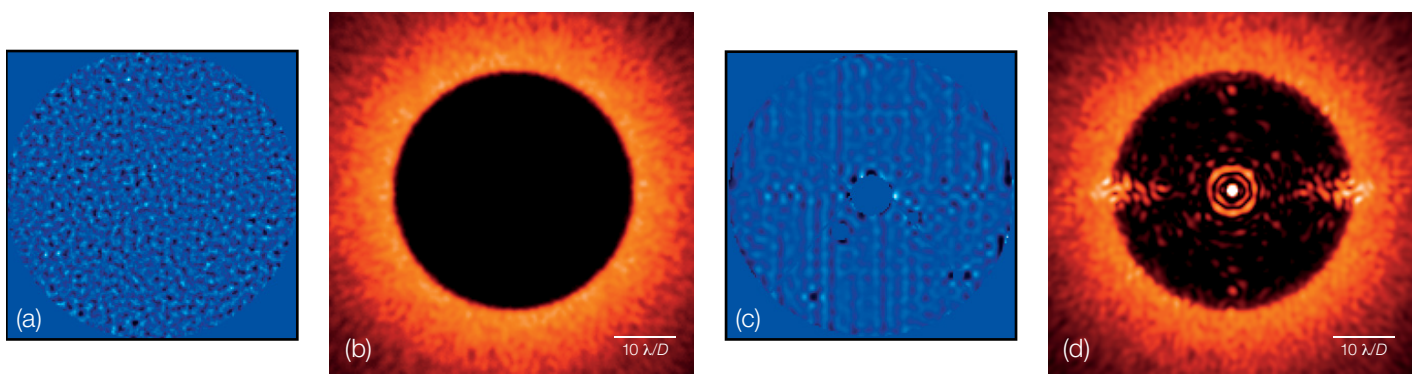

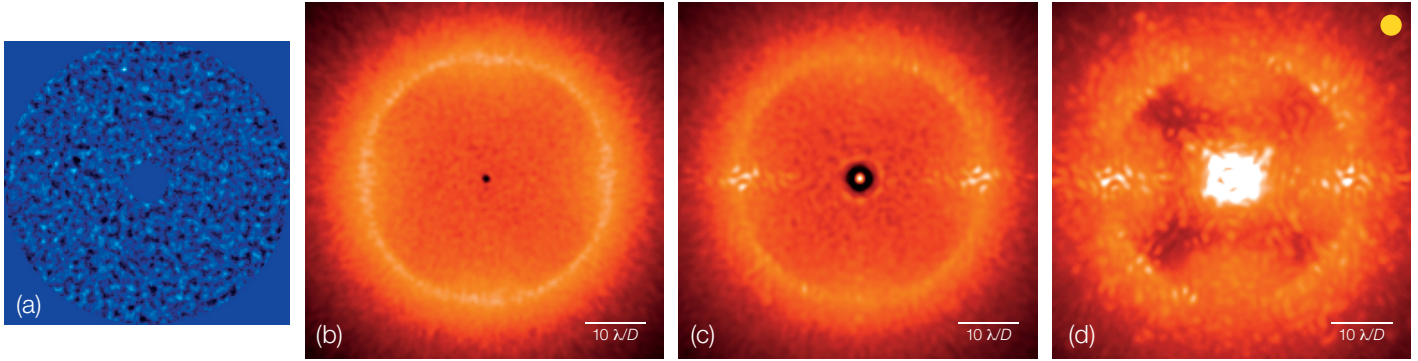

ZELDA image taken on internal source is shown in Figure 5c (Vigan et al., 2018). When propagating this phase, the resulting image shows these patterns (Figure 5d). In order to avoid the light diffracted by defective actuators of the HODM reaching the image, the coronagraph Lyot stop was remanufactured with 6 patches to hide dead actuators (Figure 2).

The aliasing error (Figure 6)
The WFS has limited spatial sampling of the incoming phase and, as a result, the uncorrected high spatial frequencies of the atmospheric turbulence may be seen by the WFS as low spatial frequencies (Figure 6a). The HODM then corrects these frequencies, but since they are not real some light is instead scattered into the corrected area (Figures 6b-d). This aliasing effect is amplified along the WFS sub-aperture directions, giving rise to a typical cross shape along this preferential direction. Moreover the aliasing effect involves spatial frequencies close to the HODM cut-off frequency and therefore the aliasing effect is more intense close to the corrected radius. To bypass this aliasing effect, a field stop (a square hole of variable size) is placed upstream of the SH-WFS to filter out the high frequencies that can neither be analysed nor corrected (Poyneer et al., 2004; Fusco et al., 2014). Depending on the seeing conditions, different filter sizes can be used to minimise aliasing; the smallest filter size can be used under very good observing conditions as this effect increases with the seeing.

The satellite spots (Figure 7)
Sometimes two perpendicular sine waves are applied to the HODM (the so-called "waffle mode", Figure 7a). This pattern creates four satellite spots in the focal plane image. Each spot is a pure copy of the star image and hence shows the same aberrations (Figures 7b and c). The intensity of the satellite spots is given by the sine wave amplitude, their position by the sine wave frequency; their direction is perpendicular to the sine wave direction. Secondary orders create multiple satellite spots in the image, which are usually too faint to be observed in SPHERE images. In addition, owing to the finite spectral bandwidth of SPHERE, the satellite spots are always slightly radially elongated (Figure 7b).

Figure 6. Illustration of the folded light in the corrected zone due to the aliasing error ($H2$-band): (a) AO residual phase showing lower spatial frequencies, (b) simulated ideal coronagraphic image, (c) simulated APLC coronagraphic image and (d) on-sky image where aliasing dominates.

This waffle mode is commonly applied at the beginning of the observing sequence to estimate the location of the centre of the star behind the coronagraph signature in the final image, which is precisely located at the intersection of the four satellite spots. Note that as a result of its manufacturing process, the grid of the HODM creates a similar pattern, provoking the presence of bright spots along the HODM grid direction (horizontally and vertically) located at 40 $\lambda/D$ in the SPHERE images ($\lambda$ being the observation wavelength and $D$ the effective telescope diameter).

Figure 7. Illustration of the satellite spots ($H2$-band): (a) waffle pattern applied on the HODM with a frequency of 14 cycles per pupil diameter, (b) simulated ideal coronagraphic image obtained with the waffle pattern added to the AO residual phase resulting in four satellite spots located at 14 $\lambda/D$, (c) simulated APLC coronagraphic image and (d) on-sky image taken with the waffle mode.

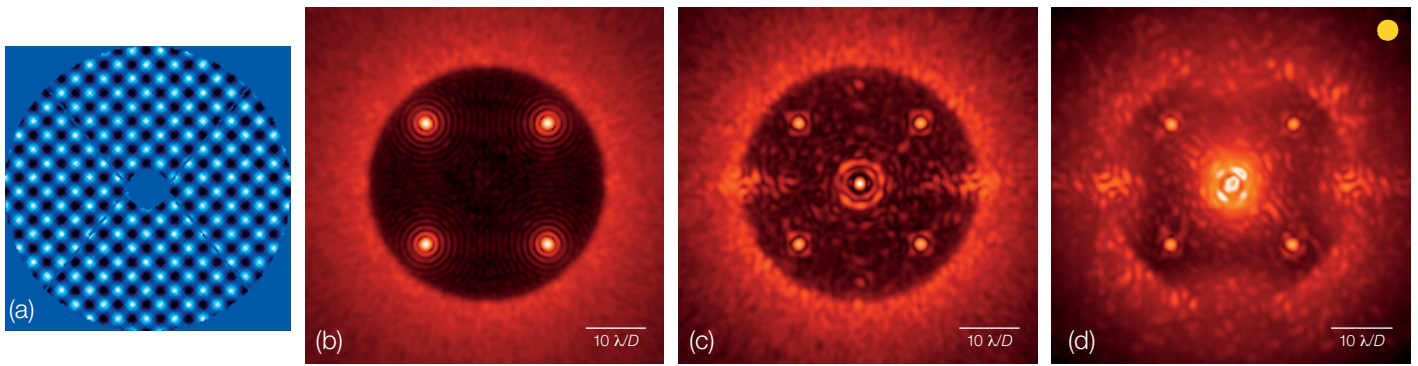

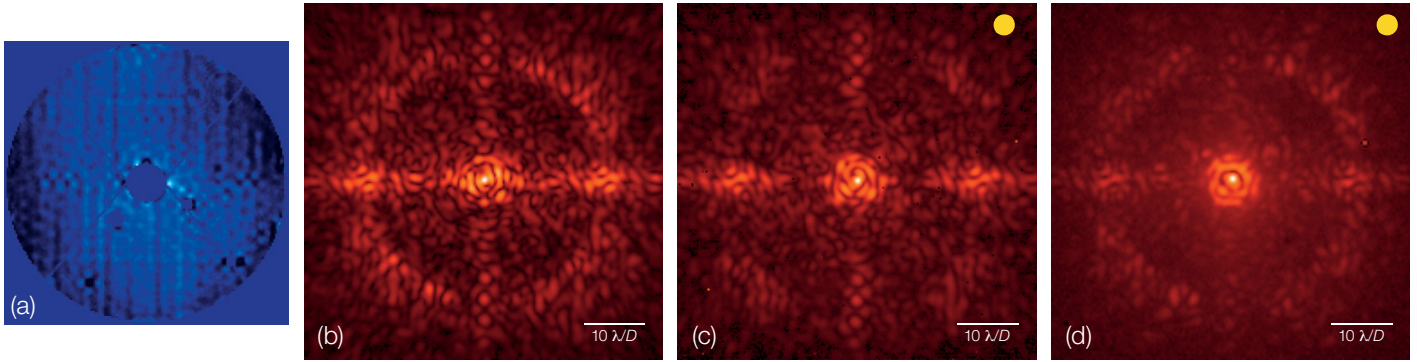

Figure 8. Illustration of the non-common path aberrations (*H2*-band): (a) non-common path aberrations phase map upstream of the coronagraph focal plane mask estimated using the ZELDA mask, (b) simulated APLC coronagraphic image using this estimated ZELDA phase map, (c) internal source image and (d) on-sky image where the non-common path aberrations are dominant.

### The contrast killers

In the context of high-contrast imaging with instruments such as SPHERE, two major aspects greatly affect the final contrast performance: (i) the errors that provoke starlight leakage out of the coronagraph; and (ii) the errors that are not temporally stable, or more generally not deterministic, and hence cannot be removed by any current post-processing techniques. In the following we focus on the errors affecting the corrected area in the images, that is to say low-order residual aberrations.

#### The non-common path aberrations (Figure 8)
Under very good conditions, current high-contrast images are limited by speckles originating from non-common path aberrations (NCPA). These are aberrations that are sensed and corrected for in the AO arm, but that are not present in the light path of the scientific subsystems and vice-versa. Like the AO residuals, they distort the wavefront so that each incoming light ray interferes with the others in the focal plane to form the "speckle field" (Figure 8b). The size of each speckle is typically that of one resolution element (1 $\lambda/D$), as for planetary signals, and their typical contrast can go up to $10^{-4}$, whereas that of the sought planetary signals is less than $10^{-6}$.

Advanced post-processing techniques are then necessary to detect exoplanet signals. NCPA that are located upstream of the coronagraph focal plane mask have been recently measured thanks to the ZELDA mask on the SPHERE internal source (Figure 8a, Vigan et al., 2018). When comparing the image simulated using this NCPA measurement (Figure 8b) to the internal source image (Figure 8c), a similar speckle field is observed. Under good observing conditions, such a speckle field is indeed limiting the contrast reached in the AO-corrected zone (Figure 8d).

#### The wind-driven halo (Figure 9)
This halo appears when high wind speeds move the upper level atmospheric turbulence across the pupil considerably faster than the AO loop can correct for. The AO residual phase shows strong atmospheric residuals with a clear directional pattern along the wind direction (Figure 9a). When propagating this phase, it produces a typical butterfly-shaped structure in the focal plane image, along the wind direction (Figures 9b–d). This temporal error significantly affects the contrast reached by the instrument (Mouillet et al., 2018). Recent studies have shown that the fast, high-altitude jet stream atmospheric layer (typically located at about 12 km above Cerro Paranal), whose wind speed can reach 50 m s$^{-1}$, is the main cause of the wind-driven halo (for example, Madurowicz et al., 2018). Moreover, this halo shows an unexpected asymmetry caused by interference between this temporal lag error and scintillation errors

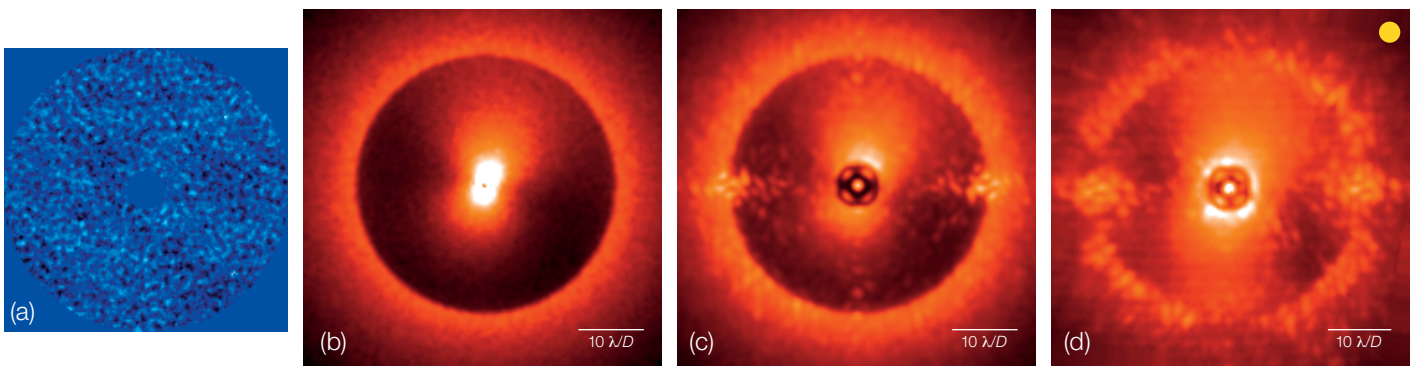

Figure 9. Illustration of the wind-driven halo due to the Jetstream layer (IFS, *Y*-band): (a) AO residual phase map showing large atmospheric residuals as ripples perpendicular to the wind direction, (b) simulated ideal coronagraphic image using only this phase map, (c) simulated APLC coronagraphic image and (d) on-sky image where the wind-driven halo dominates.

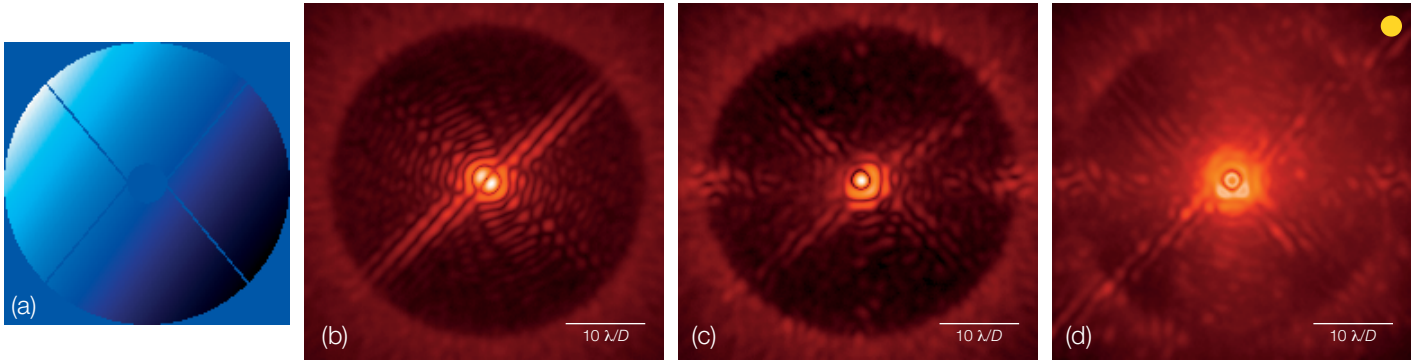

Figure 10. Illustration of the low order residuals (*K1*-band): (a) Tilt phase map added to the AO residual phase, (b) simulated ideal coronagraphic image with only tilt error added to the AO residuals, (c) simulated APLC coronagraphic image and (d) on-sky image where LORs dominate.

whose effect is also stronger with higher-altitude turbulence (Cantalloube et al., 2018).

Low-order residuals (Figure 10)
Tip-tilt errors (Figure 10a) create image jitter. Consequently, the PSF core is not correctly centred behind the coronagraph focal plane mask. In addition, the diffraction patterns from the pupil and the spiders are not entirely hidden by the Lyot stop (Figures 10b and 10c). Fast low-order residuals may arise from residual atmospheric turbulence and telescope vibrations, while atmospheric dispersion residuals and differential thermo-mechanical effects cause slow low-order residuals.

In SPHERE, these slow residuals are minimised by a differential tip-tilt sensor (Baudoz et al., 2010). This differential tip-tilt sensor uses 2% of the infrared light at the observing wavelength, picked-off just before the coronagraph focal plane mask, to estimate the position of the PSF core every second; that is then centred by the tip-tilt mirror of SPHERE. When the target star is faint (around 8 magnitudes in *H*-band) the integration time on the differential tip-tilt sensor is longer, which potentially causes stronger low-order residuals. Also, as the size of the focal plane mask is fixed, the effect of the low-order residuals in the final image is stronger when the observing wavelength increases (Figure 10d).

The low wind effect (Figure 11)
During the night, the M2 spiders can cool below the ambient air temperature by radiative losses as their emissivity is significantly higher than that of air. As a consequence, under low wind conditions, a layer of colder air — which therefore has higher refractive index — forms around the spider (Sauvage et al., 2015). When the windspeed is high, the dense air is blown away, but when the wind is slow, an abrupt change of air index is seen from one side of the spider to the other. As a result, and since the SH-WFS is insensitive to such a phase step, each quarter (or fragment) of the pupil shows a different piston, and sometimes tip-tilt phase error (Figure 11a). The corresponding PSF shape is modified, often appearing with two bright side lobes surrounding the central PSF core, and hence this is unofficially referred to as the "Mickey Mouse effect". Since the starlight is no longer concentrated in the central core (Figures 11b and 11c), this results in strong starlight leakage off the coronagraph focal plane mask (Figure 11d). This effect is always present to some degree and becomes dominant when the wind speed is too slow to reduce the temperature difference between the spiders and the ambient air (Figure 11e).

To mitigate the low wind effect at the VLT, the M2 spiders were covered with a low-emissivity coating, thus preventing strong radiative cooling. This solution has proven effective, reducing the occurrence of this effect from 18% to 3% (Milli et al., 2018).

Figure 11. Illustration of the low wind effect (*H2*-band): (a) Differential tip-tilt phase map due to low wind effect, (b) simulated non-coronagraphic PSF, (c) corresponding on-sky image of the non-coronagraphic PSF, (d) simulated APLC coronagraphic image including the differential tip-tilt phase map and (e) on-sky image where the low wind effect dominates.

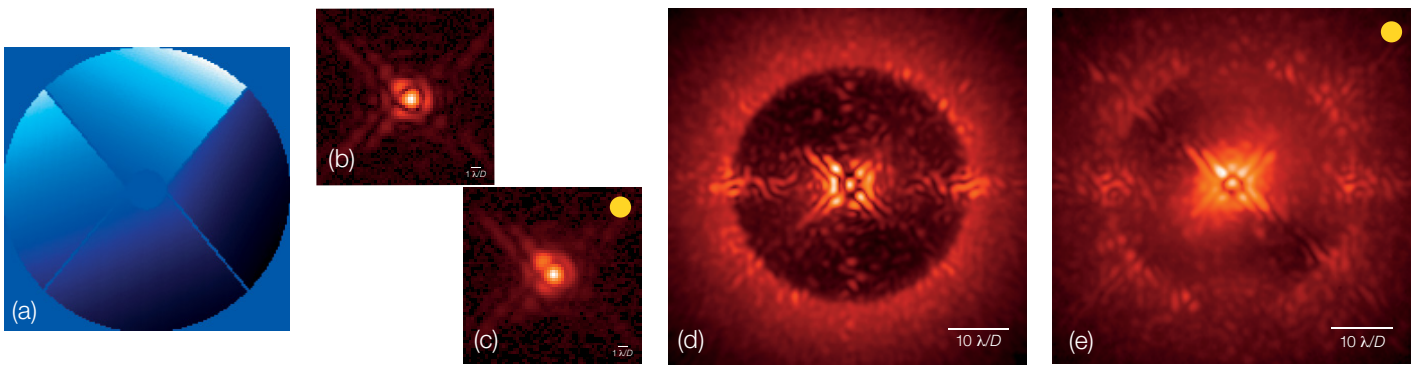

### Towards the future generation of High Contrast Imaging (HCI) instruments

Going from the first generation of exoplanet imagers, such as the Nasmyth Adaptive Optics System – Near-Infrared Imager and Spectrograph (NaCo) at ESO, to the latest generation of instruments, such as SPHERE, the contrast reached increased by an order of magnitude. This gain also revealed all the instrumental structures that are presented in this article. Analysing the origin, behaviour and effects of these structures on the current contrast performance of SPHERE offers a better understanding of high-contrast instruments. It also adds clear constraints to lead the design of future high-contrast imagers and specifically ELT instruments equipped with a high-contrast mode. Based on these considerations, SPHERE has demonstrated that scintillation, pupil fragmentation, AO temporal errors, and NCPA must be specifically tackled in the future.

In order to gain sensitivity, especially at closer separations to the star, the next generation of high-contrast instruments can build on these different aspects. For instance, on the AO side, by using a WFS less sensitive to aliasing and noise measurement (such as a Pyramid WFS; Ragazzoni et al., 1999), a faster AO loop (for example, more efficient real time computer architecture and predictive control) and going with a faster, hence more sensitive, detector and a fast deformable mirror. On the coronagraph side, by designing a coronagraph that is achromatic and less sensitive to low-order residuals (for example, pupil plane coronagraph) and offers a smaller inner working angle. On the instrument itself, correcting for the NCPA by estimating them during the observing run and by applying offsets to the HODM or by applying advanced post-processing techniques.

For ELT instruments, these different aspects will be greatly affected by the design of the telescope itself. The AO system and coronagraph will have to deal with a larger central obstruction, thicker spiders and a segmented primary mirror potentially having co-phasing errors, differential transmission and missing segments. Stay tuned for the ESO Messenger 2025 edition.

#### Acknowledgements

Faustine Cantalloube acknowledges David Mouillet and Gaël Chauvin (IPAG) for their useful comments. Arthur Vigan acknowledges support from the European Research Council (ERC) under the European Union's Horizon 2020 research and innovation programme (grant agreement No. 757561).

#### Links

[1] The SPHERE subsystems description can be found at: https://www.eso.org/sci/facilities/paranal/instruments/sphere.html
[2] The SPHERE-related press releases can be found at: https://www.eso.org/public/news/archive/search/?published_until_year=0&published_until_day=0&description=&title=&instruments=57&subject_name=&published_since_day=0&published_since_month=0&published_until_month=0&id=&published_since_year=0
[3] SPHERE filters description and transmission curves: http://www.eso.org/sci/facilities/paranal/instruments/sphere/inst/filters.html
[4] SPHERE user manual: https://www.eso.org/sci/facilities/paranal/instruments/sphere/doc/VLT-MAN-SPH-14690-0430_v100_p2.pdf

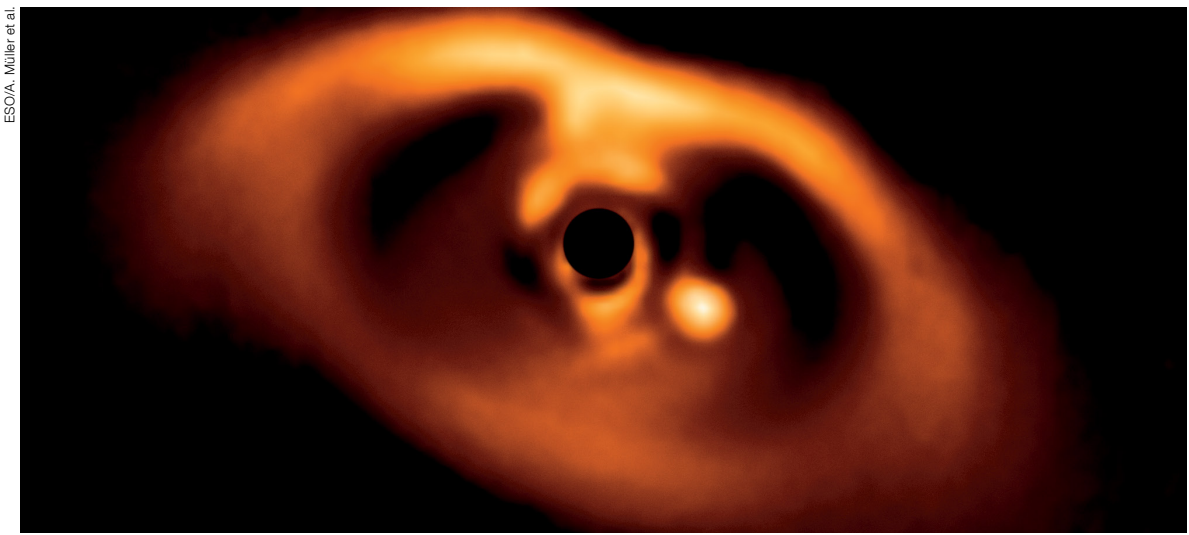

This SPHERE image of the protoplanetary disc around the young star PDS 70 reveals a planet in the act of formation.

ESO/A. Müller et al.